\documentclass[twocolumn]{aastex631} 

\usepackage{float}
\usepackage{threeparttable}

\usepackage{amsmath}
\usepackage{booktabs} 
\begin{document}

\title{Does Machine Learning Work? A Comparative Analysis of Strong Gravitational Lens Searches in the Dark Energy Survey}

\author[0000-0001-7282-3864]{J. González}\affiliation{Physics Department, University of Wisconsin-Madison, 1150 University Avenue Madison, WI 53706, USA} \affiliation{NSF-Simons AI Institute for the Sky (SkAI), 172 E. Chestnut St., Chicago, IL 60611, USA} \affiliation{Center for Interdisciplinary Exploration and Research in Astrophysics (CIERA), Northwestern University, 1800 Sherman Ave, Evanston, IL 60201, USA \\ \href{mailto:jimenagonzalez@northwestern.edu}{jimenagonzalez@northwestern.edu}}
\author[0000-0001-5564-3140]{T. Collett}\affiliation{Institute of Cosmology and Gravitation, University of Portsmouth, Burnaby Rd, Portsmouth PO1 3FX, UK}
\author[0000-0003-1391-6854]{K. Rojas}\affiliation{University of Applied Sciences and Arts of Northwestern Switzerland, School of Engineering, 5210 Windisch, Switzerland} \affiliation{Institute of Cosmology and Gravitation, University of Portsmouth, Burnaby Rd, Portsmouth PO1 3FX, UK}
\author[0000-0001-8156-0429]{K. Bechtol} \affiliation{Physics Department, University of Wisconsin-Madison, 1150 University Avenue Madison, WI 53706, USA}
\author[0000-0002-9654-1711]{J. A. Acevedo Barroso} \affiliation{Institute of Physics, Laboratory of Astrophysics, Ecole Polytechnique F\'ed\'erale de Lausanne (EPFL), Observatoire de Sauverny, 1290 Versoix, Switzerland}
\author[0000-0002-6449-3970]{A. Melo}\affiliation{Max-Planck-Institut für Astrophysik, Karl-Schwarzschild-Str. 1, 85748 Garching, Germany} \affiliation{Technical University of Munich, TUM School of Natural Sciences, Physics Department, James-Franck-Straße 1, 85748 Garching, Germany}
\author[0000-0001-7714-7076]{A. More} \affiliation{The Inter-University Centre for Astronomy and Astrophysics (IUCAA), Post Bag 4, Ganeshkhind, Pune 411007, India}\affiliation{Kavli Institute for the Physics and Mathematics of the Universe (IPMU), 5-1-5 Kashiwanoha, Kashiwa-shi, Chiba 277-8583, Japan}
\author[0000-0001-6116-2095]{D. Sluse} \affiliation{STAR Institute, University of Li{\`e}ge, Quartier Agora, All\'ee du six Ao\^ut 19c, 4000 Li\`ege, Belgium}
\author[0000-0001-7958-6531]{C. Tortora} \affiliation{INAF -- Osservatorio Astronomico di Capodimonte, Salita Moiariello 16, I-80131, Napoli, Italy}
\author[0009-0002-8896-6100]{P. Holloway} \affiliation{University of Oxford (Sub-department of Astrophysics, University of Oxford, Denys Wilkinson Building, Keble Road, Oxford, OX1 3RH, UK)} \affiliation{Institute of Cosmology and Gravitation, University of Portsmouth, Burnaby Rd, Portsmouth PO1 3FX, UK}
\author[0009-0004-7751-1914]{N. E. P. Lines} \affiliation{Institute of Cosmology and Gravitation, University of Portsmouth, Burnaby Rd, Portsmouth PO1 3FX, UK}
\author[0000-0002-0730-0781]{A. Verma} \affiliation{University of Oxford (Sub-department of Astrophysics, University of Oxford, Denys Wilkinson Building, Keble Road, Oxford, OX1 3RH, UK)}


\begin{abstract}

We present a systematic comparison of three independent machine learning (ML)–based searches for strong gravitational lenses applied to the Dark Energy Survey \citep{Jacobs_2019_high, Jacobs_2019_extended, Rojas_2022, gonzalez2025}. Each search employs a distinct ML architecture and training strategy, allowing us to evaluate their relative performance, completeness, and complementarity. Using a visually inspected sample of 1651 systems previously reported as lens candidates, we assess how each model scores these systems and quantify their agreement with expert classifications. The three models show progressive improvement in performance, with F1-scores of 0.31, 0.35, and 0.54 for Jacobs, Rojas, and González, respectively. Their completeness for moderate- to high-confidence lens candidates follows a similar trend (31\%, 52\%, and 70\%). When combined, the models recover 82\% of all such systems, highlighting their strong complementarity. Additionally, we explore ensemble strategies: average, median, linear regression, decision trees, random forests, and an Independent Bayesian method. We find that all but averaging achieve higher maximum F1 scores than the best individual model, with some ensemble methods improving precision by up to a factor of six. These results demonstrate that combining multiple, diverse ML classifiers can substantially improve the completeness of lens samples while drastically reducing false positives, offering practical guidance for optimizing future ML‑based strong lens searches in wide‑field surveys.
\end{abstract}



\section{Introduction} \label{sec:introduction}

Strong gravitational lensing is the phenomenon in which the light from an astronomical source is deflected by the gravitational potential of a massive foreground object. This effect produces multiple magnified images of the source, and its observables are sensitive to the mass distribution of the deflector and cosmological parameters. As a result, strong lensing is a powerful probe for studying the evolution and structure of elliptical galaxies and the nature of dark matter and dark energy.

Since strong gravitational lens systems are exceptionally rare, identifying them in large datasets is challenging. Early targeted searches involved manually inspecting large numbers of high-resolution images from small-area high-resolution astronomical surveys, typically selected using modest preselection criteria (e.g., \citealp{Fassnacht_2004, Fassnacht_2006, Moustakas_2007, Faure_2008, Jackson_2008, Newton_2009, More_2011}). However, wide-field astronomical surveys require automated search techniques to process and filter vast amounts of data efficiently. An early example of such a technique is RingFinder, an algorithm designed to detect blue residuals around smooth red galaxy light profiles \citep{alard2006automateddetectiongravitationalarcs}. Other approaches have focused on identifying arc-like or elongated features in astronomical images (e.g., \citealp{More_2012, Gavazzi_2014}). Another example is \texttt{YATTALENS} \citep{Sonnenfeld_2017}, which identifies galaxy-scale lenses by modeling the systems and subtracting the deflector's light from the image. Additional search efforts have implemented algorithms that preselect image cutouts containing objects with the characteristic blue and red color features of strong gravitational lenses for subsequent visual inspection (e.g., \citealp{Diehl_2017, ODonell_2022}). While these automated approaches still rely heavily on human inspection, they are particularly effective at recovering systems with smaller Einstein radii. Moreover, the candidates identified through such methods are valuable for constructing training and testing samples for machine learning (ML)–based searches.

State-of-the-art search methods employ ML techniques to pre-select strong lensing candidates, significantly outperforming previous search methods \citep{Metcalf_2019_SL_challenge}. In the past few years, numerous ML-based searches have been applied to multiple astronomical surveys, including the Hyper Supreme Camera  (HSC, \citealp{Ca_ameras_2021, Shu_2022, jaelani2023surveygravitationallylensedobjects}), the Kilo-Degree Survey (KiDS, \citealp{Petrillo_2017, Petrillo_2019, Li_2020, Li_2021, Grespan_2024}), the Dark Energy Spectroscopic Instrument DECam Legacy Survey (DESI-LS, \citealp{Huang_2020, Huang_2021, Stein_2022, Dawes_2023, Storfer_2024}), the DECam Local Volume Exploration Survey (DELVE, \citealp{Zaborowski_2023}), the Panoramic Survey Telescope and Rapid Response System Survey (Pan-STARRS, \citealp{Ca_ameras_2020}), the Ultraviolet Near Infrared Optical Northern Survey (UNIONS, \citealp{Savary_2022, barroso2025}), and others.

Despite the widespread use of ML techniques in strong lensing searches, key gaps remain. First, the extent to which ML models fail to identify genuine strong gravitational lenses has not been thoroughly quantified. In most ML-based searches, only high-scoring images are selected for visual inspection, leaving the number of true lenses missed due to low ML-assigned scores largely unknown. A first attempt to address this was made by the \citet{euclid_walmsley}, where citizen scientists visually inspected a random sample of 40000 targets that had not been highly scored by ML models. They estimated a missed-lens rate of 0.79 lenses per thousand targets, highlighting that low-scoring subsets may contain real lenses. Understanding and quantifying ML completeness is also crucial for optimizing score thresholds and combining predictions from multiple models to improve recovery rates while minimizing false positives. Second, only one study to date has systematically compared the performance of different ML models on a shared testing sample \citep{more2024systematiccomparisonneuralnetworks}. Further analyses are needed using large, realistic testing sets that capture the complexity and diversity of modern survey data. Addressing these gaps is essential for identifying the limitations of current ML-based searches and enhancing their reliability in future applications.

The Dark Energy Survey (DES) is an astronomical survey that utilizes the Dark Energy Camera (DECam, \citealp{Flaugher_2015}), mounted on the Victor M. Blanco Telescope at the Cerro Tololo Inter-American Observatory in Chile. Due to its wide-field coverage of \raisebox{0.5ex}{\texttildelow}5000 deg² of the southern sky, multiple strong lensing searches have been applied to its data (e.g. \citealp{Nord_2016, Diehl_2017, Jacobs_2019_high, Jacobs_2019_extended, Nord_2020, ODonell_2022, Rojas_2022, gonzalez2025}). Notably, at least three of these searches have employed ML techniques \citep{Jacobs_2019_high, Jacobs_2019_extended,  Rojas_2022, gonzalez2025}, each identifying hundreds of strong lensing candidates. The methodologies of these three searches were developed independently, using different network architectures and training data sets. Since the three ML algorithms were applied to large volumes of DES data, they provide an opportunity to assess and compare their performance and completeness. 


In addition, these independently developed ML classifiers enable the exploration of ensemble strategies that combine their outputs into a unified, more robust model. Ensemble methods tend to outperform individual ML models by improving generalization, increasing robustness, and reducing variance \citep{Ganaie_2022}. In the context of strong lensing searches, several ensemble approaches have been explored. Most studies have employed ``decision fusion" techniques, in which the outputs of individual models are combined according to predefined rules or heuristics. For instance, \cite{Nagam_2023} and \cite{Andika_2023} averaged the outputs of multiple ML models, while the best-performing ensemble method in \cite{Holloway_2024} used a Bayesian Independent approach that integrates ML predictions with citizen science scores. The latter methodology was assessed with Euclid data, achieving a sample purity of $52 \pm 2\%$ at 50\% completeness \citep{euclid_holloway}. Another ensemble approach, known as ``stacking", involves using the outputs of individual models as input to train a new ML model \citep{Akhazhanov_2022}.

Most recently, the Euclid Collaboration conducted a search for strong gravitational lenses in the Euclid Quick Data Release Q1 using multiple independently developed ML models \citep{euclid_lines}. A total of five ML models were used, each featuring a different network architecture and trained on partially overlapping but distinct training datasets. To select images for subsequent visual inspection, experts first estimated the performance of each method by inspecting the top 1000 ranked images of each. Larger samples were then shown to citizen scientists, with priority given to the models that demonstrated superior performance according to the expert evaluation \citep{euclid_walmsley}. The top 10000 ranked images from each model were included, along with an additional 10000 images from each of the two best-performing ML models. From the next 10000 ranked images of the remaining three ML models, a random subset of 5000 was also selected.

The purpose of this work is threefold: (i) to perform a comparative analysis of the methods developed by
\citet[hereafter \textit{Jacobs}]{Jacobs_2019_high,Jacobs_2019_extended},
\citet[hereafter \textit{Rojas}]{Rojas_2022}, and
\citet[hereafter \textit{Gonz\'alez}]{gonzalez2025};
(ii) to analyze the distribution of ML scores assigned to strong lensing candidates and therefore determine if ML techniques are effective at identifying them; and
(iii) to evaluate different ensemble techniques for combining individual ML outputs. To achieve this, we first identify the sample of input images processed by the three search methods and select systems reported in the SLED database\footnote{\url{https://sled.amnh.org/}} \citep{Vernardos2024} as strong lensing candidates. The resulting sample, consisting of 1651 images, is visually inspected by experts assigning a score representing the confidence that each candidate is a genuine strong lensing system. These scores are then used to classify candidates into different confidence categories that are used to calculate performance metrics.

This paper is organized as follows: Section~\ref{Sec:Summary_searches} provides an overview of the network architectures and training samples used in each of the three search efforts. In Section~\ref{Sec:Methodology}, we describe the selection process of the targets analyzed by the ML algorithms, the identification of strong lensing candidates for expert visual inspection, and the DES cutout images used in this process. This section also outlines the procedure experts followed to quantify the confidence of the candidates. Section~\ref{Sec:Results} presents our analysis and results, which include a comparison of the performance of the three ML models and an assessment of their completeness and complementarity. Section~\ref{sec:Ensemble_methods} outlines the ensemble techniques we investigate in this work and compares their performance. Finally, Section~\ref{Sec:Conclusions} summarizes our conclusions and discusses their implications for future ML-based strong lensing searches.

\section{Summary of the search methods} \label{Sec:Summary_searches}

\subsection{Jacobs}

\cite{Jacobs_2019_extended, Jacobs_2019_high} employs a Convolutional Neural Network (CNN) with approximately 9 million trainable parameters, similar in complexity to AlexNet \citep{Alexnet}. The architecture consists of four convolutional layers with kernel sizes of 11, 5, 3, and 3, followed by two fully connected layers, each containing 1024 neurons. Each convolutional layer uses a ReLU activation function \citep{reul_activation} and is followed by a 2$\times$2 max pooling layer. This team selected this network architecture due to its high accuracy during training and its strong performance, with a similar network architecture ranking third in the Bologna Lens Finding Challenge \citep{Metcalf_2019_SL_challenge}. During training, a binary cross-entropy loss was minimized, and training was halted if the validation loss did not improve by more than \(10^{-4}\) for six epochs.

This training sample consisted of 200,000 images in the $g$-, $r$-, and $i$-bands, equally divided into positive and negative examples. Positive examples are simulated using the ``redMaGiC" catalog  \citep{redmagic}, which consists of luminous red galaxies (LRGs), to act as the galaxy deflectors. The photometric redshift of the deflector galaxy is extracted from the DES catalog TABLE \citep{Abbott_2018_datarelease_1}, and its velocity dispersion value is calculated with the Hyde \& Bernardi fundamental plane relation \citep{fundamental_plane}. For each galaxy in the catalog, the team created three simulations placing sources at different positions in the plane. The deflector’s mass profile is modeled as a Singular Isothermal Ellipsoid (SIE), its light distribution follows an elliptical de Vaucouleurs profile, and the source's light profile is modeled as an exponential disk. The lensed source images are generated with the GRAVLENS code \citep{keeton2001computationalmethodsgravitationallensing} and they are combined with real DES images of the deflector to obtain the complete simulation. Negative training examples are randomly drawn from their target search catalog, as this sample is very unlikely to contain strong lenses. The selection criteria for this catalog are described in more detail in Section~\ref{subsec:data_selection}.

\subsection{Rojas}

\cite{Rojas_2022} utilizes EfficientNet, a family of CNN architectures designed to balance accuracy and computational efficiency \citep{tan2020efficientnetrethinkingmodelscaling}. Among these, EfficientNet B0 is the simplest and smallest model, serving as the baseline for the family. Its architecture employs compound scaling, a method that systematically adjusts the network's depth, width, and resolution to optimize performance while minimizing computational cost. Despite its simplicity, EfficientNet B0 achieves high performance in image classification tasks, making it well suited for applications with limited resources. This work uses an EfficientNet B0 ML model with around 4 million trainable weights. During the training process, a binary cross-entropy loss function was minimized using an Adam optimizer \citep{kingma2017adammethodstochasticoptimization}.

This training sample comprised 200,000 images in the $g$-, $r$-, and $i$-bands, equally divided into positive and negative examples. To generate positive examples, real images of high-redshift galaxies observed by the Hubble Space Telescope (HST), combined with color information from HSC, are used as source galaxies, allowing a realistic morphology and color distribution. LRGs serve as deflector galaxies, with their redshifts and velocity dispersions estimated using a K-nearest-neighbors algorithm based on the best magnitude matches. This algorithm was trained using as input the $gri$ magnitudes, redshifts, and velocity dispersions of about a million Sloan Digital Sky Survey (SDSS) galaxies. The deflectors are modeled with an SIE mass profile and a Sersic light profile. The Einstein radius values are drawn from a uniform distribution between 1.2 and 3 arcseconds. The complete simulation combines a real DES image of the deflector with a lensed HST/HSC source image created using the {\tt Lenstronomy}\footnote{\url{https://github.com/sibirrer/lenstronomy}} \citep{birrer2018lenstronomymultipurposegravitationallens, Birrer2021_lenstronomy} Python package. To enhance the visibility of lensing features, the source's original brightness is artificially increased by one magnitude, and it is positioned close to the caustic curves to increase its magnification. Negative examples are extracted from images of LRGs that are not used in the positive training images.

\subsection{González} \label{subsec:method_gonzalez}

\cite{gonzalez2025} implements a Vision Transformer (ViT, \citealp{vision_transformer}) as its machine learning model. The ViT architecture uses a transformer encoder, originally designed for natural language processing, but adapted for image processing by dividing input images into patches and treating each patch as a token. Unlike CNNs, which initially focus on local features, the ViT analyzes the entire image from the start, allowing it to capture long-range dependencies more effectively. Within the ViT architecture family, this team selected the ViT-Base/16 model, which partitions images into 16×16 pixel patches and contains approximately 86 million trainable parameters, making it the smallest model in the series. The chosen model is pre-trained on the ImageNet 21k dataset \citep{imagenet}, which contains over 14 million images across approximately 21000 categories. Pre-training on this large and diverse dataset allows the model to learn a broad range of general visual features, which enhances its ability to generalize to new data and improves its performance on classification tasks. During training, a cross-entropy loss function is minimized using the Adam optimizer.

The training sample in this work was constructed through an iterative process, where a trained ML model was applied to random images from the target search catalog, and then, based on the model's performance on this subset, the training set was refined. This approach, known as Interactive Machine Learning, is a ``Human-in-the-Loop" technique that involves close interaction between the ML model and the developers to improve the performance of the model. In addition, this ML search was framed as a multi-class classification task with nine different categories. The total training sample consisted of \raisebox{0.5ex}{\texttildelow}40000 images in the $g$-, $r$-, and $i$-bands, with the largest classes being the positive class (strong lenses) and ``Red Spheroids," containing 14000 and 15000 images, respectively. The remaining seven classes, which included common false positives such as ring galaxies, spiral galaxies, galaxies with diffuse or extended structures, and galaxies with edge-on companions, contained between 1000 and 2000 examples each. The positive examples consisted of simulated strong gravitational lenses, in which real DES images of the deflector galaxy were used. The deflector follows an SIE mass profile, and a Sersic light profile models both the deflector and the source. The Einstein radii of the simulations span a range of approximately $1^{\prime\prime}$ to $6^{\prime\prime}$, with most systems falling between $1.7^{\prime\prime}$ and $4.2^{\prime\prime}$. To ensure clear lensing features, simulation properties were constrained, and the brightness of the lensed sources was artificially increased by two magnitudes, making the simulations more obvious and less realistic.

\section{Methodology}\label{Sec:Methodology}

\subsection{Data Selection}\label{subsec:data_selection}

First, we identify the images analyzed as input by the three ML models from Jacobs, Rojas, and González. Each of these works adopted different criteria to create their target search catalog of images. \cite{Jacobs_2019_high, Jacobs_2019_extended} applied a color selection cut of $0 < g - i < 3$ and $-0.2 < g - r < 1.75$, resulting in a sample of 7.9 million targets. \cite{Rojas_2022} adopted a similar color selection cut of $1.8 < g - i < 5$ and $0.6 < g - r < 3$, in addition to a magnitude (\texttt{mag\_auto} column of the first DES data release, \citealp{Abbott_2018_datarelease_1}) cut of $18 < r < 22.5$, $g > 20$ and $i > 18.2$, yielding a sample of 18.7 million targets. \cite{gonzalez2025} did not apply color selection cuts but selected targets likely to be extended objects (\texttt{ext\_coadd}~$>~1$) that passed a magnitude (\texttt{mag\_aper\_8}) selection cut of $15 < i < 23.5$, using data from the second DES data release \citep{Abbott_2021_datarelease_2}, resulting in a sample of 236 million targets. By cross-matching the three target search catalogs using a $1^{\prime\prime}$ positional tolerance, we recover a set of approximately 750,000 targets that are common to all samples, which we refer to as the ``Intersection" sample. 

Given the large size of each target search catalog, we rounded the ML scores to four decimal places to reduce file sizes and computational overhead. This discretization significantly reduced the number of unique ML score values, yielding approximately 6,000–10,000 distinct values in the Intersection sample for each model. Since raw ML scores are typically uncalibrated and their distributions can vary significantly across models, we transformed the scores into normalized ranks. This approach is particularly suitable in the context of strong lensing searches, where the focus lies on the top-ranked candidates rather than the absolute score values. Ranks were computed within the Intersection sample using the ``dense" method from the \textit{Pandas} module \texttt{rank} function, which assigns the same rank to repeated values without skipping subsequent positions. The resulting ranks were then normalized by dividing by the highest rank value.

\begin{figure}[htbp]
 \centering
 \includegraphics[width=1\columnwidth]{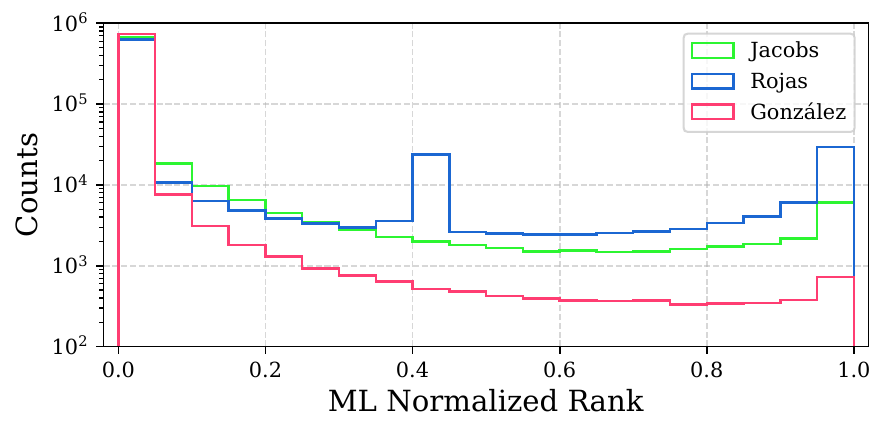}
\caption{Distribution of normalized ranks assigned by each search effort for all astronomical targets processed by the three ML models (Intersection sample). The histogram illustrates differences in score distributions.}
\label{Fig:Histo_ML_scores}
\end{figure}

Figure~\ref{Fig:Histo_ML_scores} shows the distributions of normalized ranks assigned by each ML model to the Intersection sample. The distributions are similar across the three models: the vast majority of targets received scores close to zero, while a secondary peak appears at high scores. The Rojas model shows a peak in the middle of the histogram, where many targets receive an identical ML raw score of 0.43. This feature may result from factors such as regularization, architectural constraints, among others. A detailed investigation is beyond the scope of this work. 

Within the Intersection sample, 1651 systems have been reported as strong lensing candidates in the SLED database \citep{Vernardos2024}. These systems were identified in at least 42 different publications based on various astronomical surveys, including DES, DESI-LS, HSC, KiDS, UNIONS, among others. Appendix~\ref{app:sled} lists the specific references and the number of systems from each that overlap with our sample of 1651 systems. The SLED database also provides a score, here referred to as the ``SLED score", ranging from 0 to 3, which reflects the confidence that a system is a genuine strong lens. At the time of this analysis, the majority of SLED scores were derived from assessments made by the authors of the original publications in which the systems were identified. 


\begin{figure*}[htbp]
 \centering
 \includegraphics[width=\textwidth]{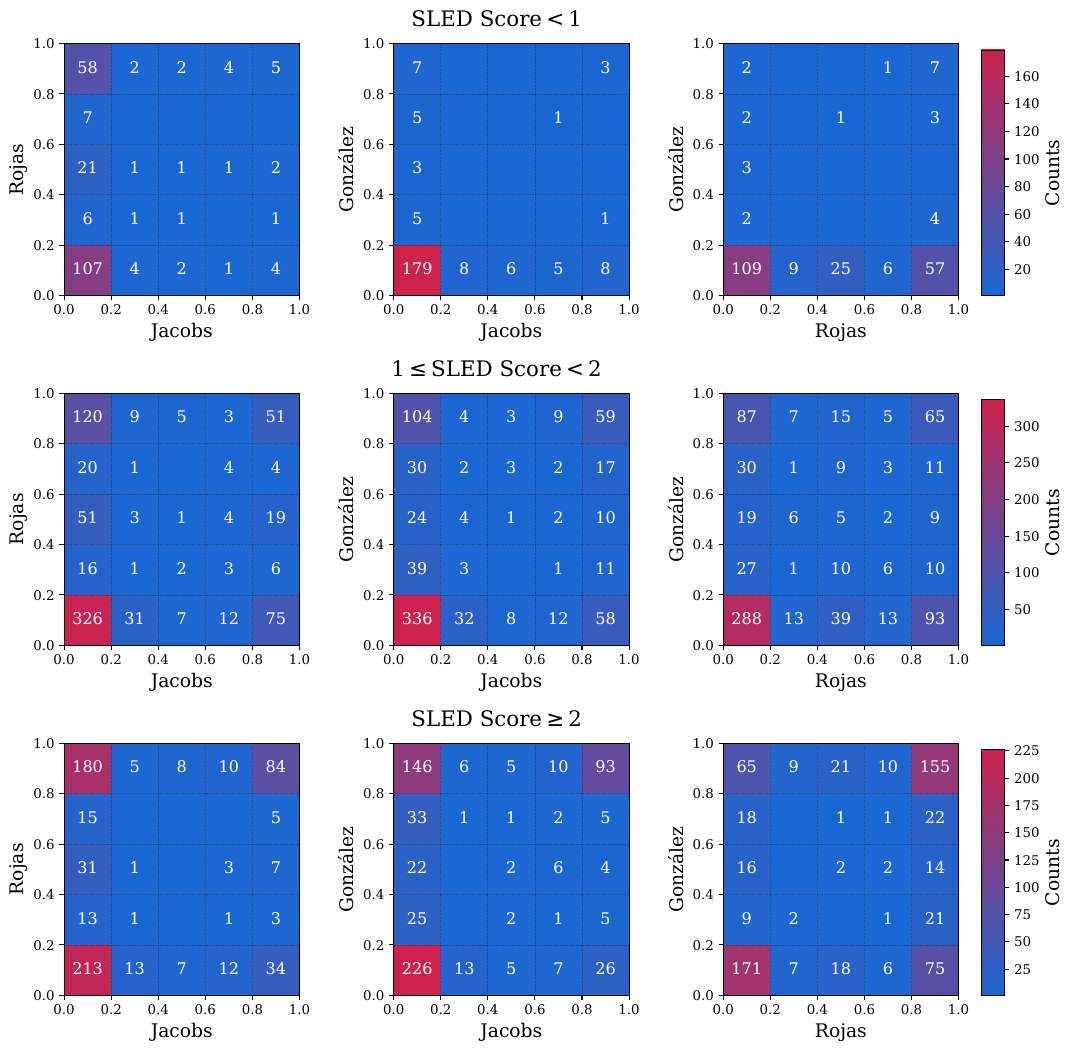}
\caption{2D histograms comparing the ML normalized ranks from the three ML models for systems reported as strong lensing candidates in the SLED database. Each row corresponds to a different range of SLED score, which indicates the confidence level of the candidates (0 to 3). This visualization examines how ML models score candidates with varying confidence levels.}
\label{Fig:2d_Histo_SLED}
\end{figure*}

As an initial approach to study the completeness of the three search methodologies, we present Figure~\ref{Fig:2d_Histo_SLED}, which shows 2D histograms of the ML normalized ranks for the 1651 systems previously reported as strong lensing candidates, grouped by different SLED score ranges. This figure reveals that a large fraction of the intermediate and high-confidence candidates receive low scores from the three ML models. For instance, among the systems with SLED scores greater than 2, 70\%, 43\%, and 43\% received ML normalized ranks below 0.2 from Jacobs, Rojas, and González, respectively. This effect likely arises because the SLED candidates come from multiple astronomical surveys with differing imaging characteristics and were identified by different research teams. As a result, the scores are heterogeneous and may not reflect the confidence of a candidate based on DES imaging specifically. For this reason, it is necessary to produce a homogeneous set of scores based on DES imaging through visual inspection by a team of experts.

\subsection{DES Cutout Images}

Both \cite{Jacobs_2019_high, Jacobs_2019_extended} and \cite{Rojas_2022} applied their ML search algorithms to coadded images from the first DES data release \citep{Abbott_2018_datarelease_1}, while \cite{gonzalez2025} used the second data release \citep{Abbott_2021_datarelease_2}. Although both datasets cover the entire DES footprint, the second release is significantly deeper, with a median coadded catalog depth for a $1.95^{\prime\prime}$ diameter aperture at a signal-to-noise ratio=10 of $g$=24.7, $r$=24.4, and $i$=23.8 \citep{Abbott_2021_datarelease_2}, compared to $g$=24.3, $r$=24.1, and $i$=23.4 from the first release \citep{Abbott_2018_datarelease_1}. Additionally, each team adopted different image cutout sizes: $26.4^{\prime\prime}$ for Jacobs, $13.2^{\prime\prime}$ for Rojas, and $11.9^{\prime\prime}$ for González. In this work, the coadded images for visual inspection are extracted from the second DES data release, and we adopt an intermediate cutout size of $19.8^{\prime\prime}$.

\subsection{Visual Inspection}

\begin{figure*}[htbp]
 \centering
 \includegraphics[width=0.8\textwidth]{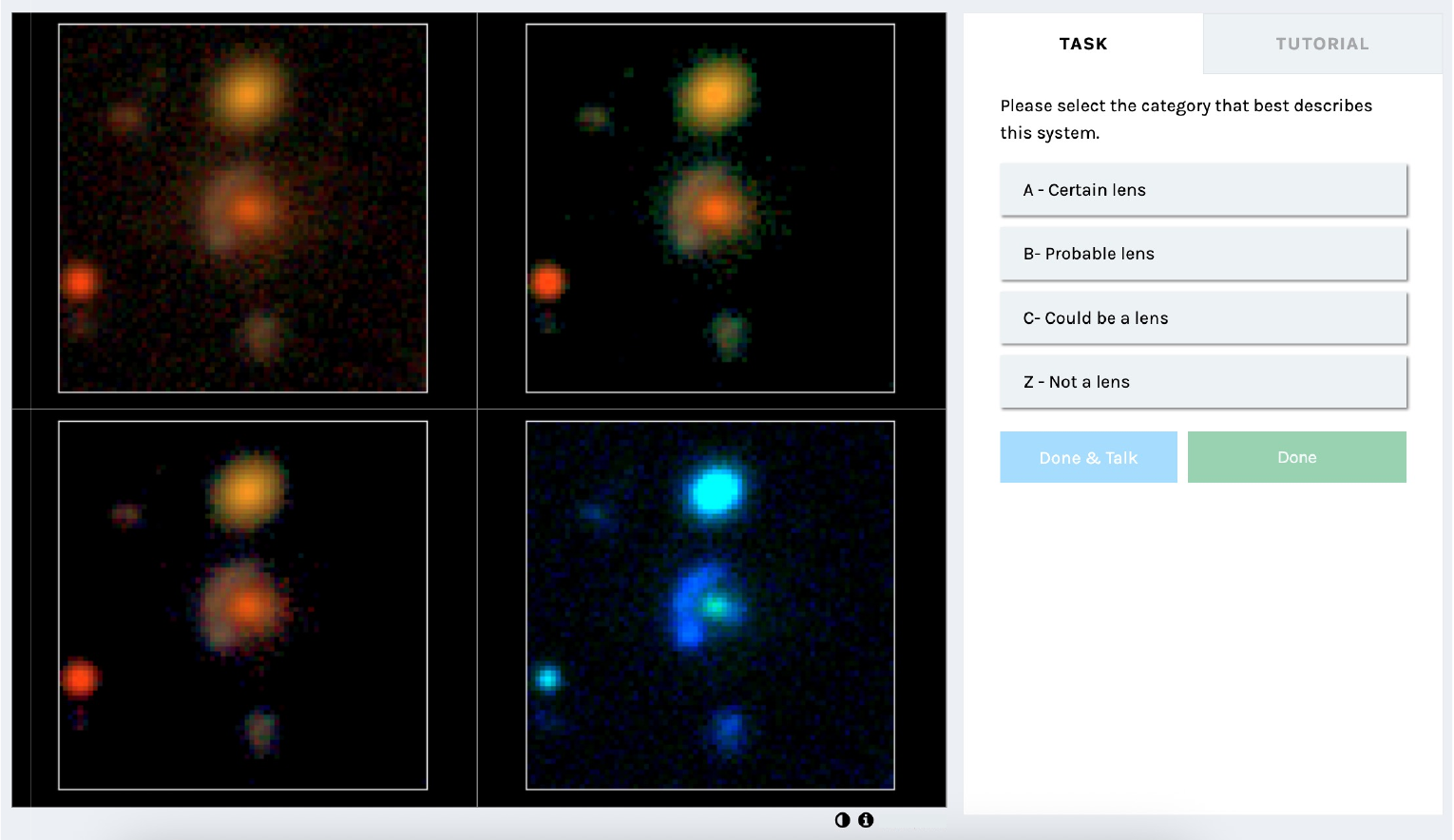}
\caption{Screenshot from the visual inspection project on Zooniverse. Strong lensing experts were shown each system in four distinct PNG settings designed to highlight different image features. Experts were asked to categorize each system into four classes: A-certain lens, B-probable lens, C-could be a lens and Z-not a lens.}
\label{Fig:Zooniverse_panel}
\end{figure*}

A team of seven experts inspected the 1651 systems reported as strong lensing candidates in the SLED database. The visual inspection was conducted on the Zooniverse platform, where each system was displayed with four different PNG settings designed to highlight various color features as shown on Figure~\ref{Fig:Zooniverse_panel}. Three of these settings were generated using different input parameters in the \texttt{make\_lupton\_rgb} function from the \textit{Astropy} Python package \citep{astropy_2013, astropy_2018}. The image in the bottom right of the panel was generated by strongly suppressing the $r$-band signal while enhancing the $g$- and $b$-band signals by factors of 3 and 15, respectively. To further emphasize image features, the contrast of the resulting image was increased by amplifying the difference between its bright and dark regions.

\begin{figure}[htbp]
 \centering
 \includegraphics[width=\columnwidth]{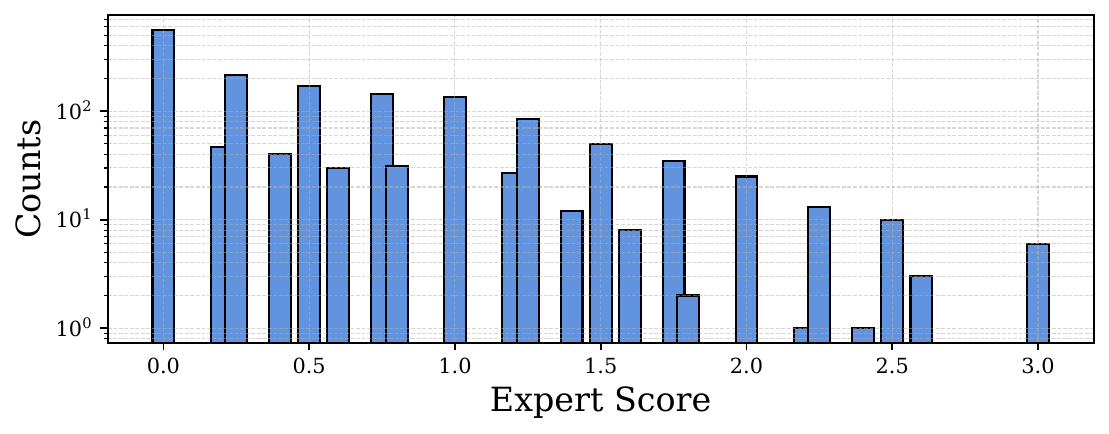}
\caption{Histogram of the Expert Scores assigned to systems previously reported in the SLED database as strong lensing candidates. The scores were determined through visual inspection of images from the second DES data release. To reflect the discrete nature of the Expert Scores, each bar corresponds to a unique score value rather than to a continuous histogram bin.}
\label{Fig:Histo_Expert_scores}
\end{figure}

Experts were asked to classify each system as one of the following: A-certain lens, B-probable lens, C-could be a lens, or Z-not a lens. Then, these categories were assigned numerical values (A: 3, B: 2, C: 1, Z: 0). While the classifications are generally reliable, occasional misclassifications can occur due to the fast-paced nature of the task and to Zooniverse not allowing users to revise previous responses. To account for this, we aggregate the scores of each system by discarding the highest and lowest values and averaging the remaining ones. This approach helps reduce the impact of accidental errors while yielding a continuous score that facilitates threshold selection. Figure~\ref{Fig:Histo_Expert_scores} shows the distribution of calibrated scores, which we refer to as the ``Expert Score".

\begin{figure*}[htbp]
 \centering
 \includegraphics[width=\textwidth]{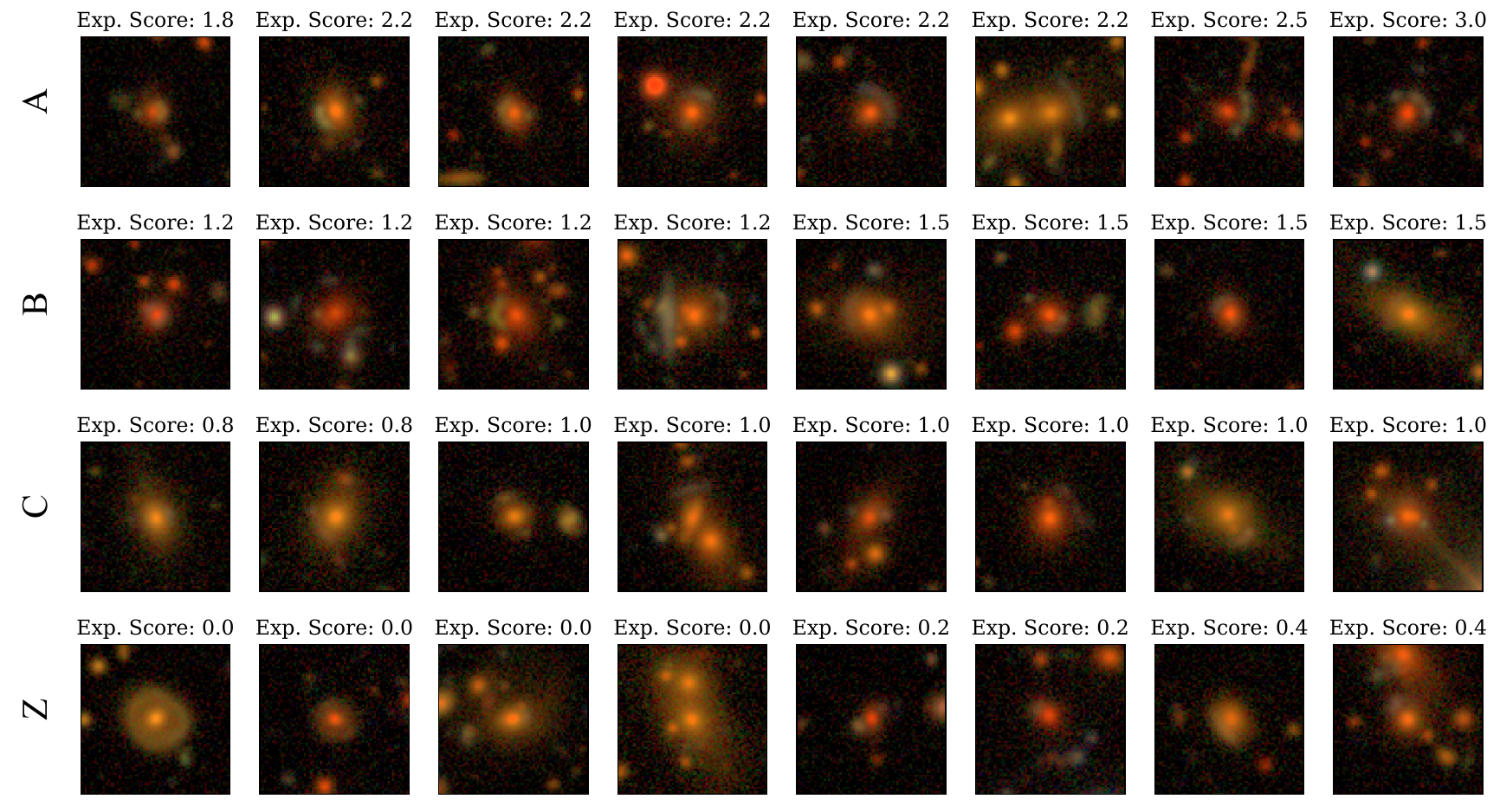}
\caption{Random selection of eight systems from each category (A, B, C, Z). Within each row, systems are ordered from left to right by increasing Expert Score. This figure provides an intuitive visual sense of the typical characteristics within each category.}
\label{Fig:categories_lenses}
\end{figure*}

To evaluate the performance of the ML models, we classify the inspected systems based on their Expert Score into four categories reflecting the level of confidence in their strong lensing nature: A–definite lenses, B–probable lenses with some ambiguity, C–ambiguous cases, and Z–systems unlikely to be lenses. These categories are defined using the following Expert Score thresholds:
\begin{itemize}
    \item A: Expert Score $\geq$ 1.8, containing 61 systems
    \item B: 1.2 $\leq$ Expert Score $<$~1.8, containing 217 systems
    \item C: 0.8 $\leq$ Expert Score $<$~1.2, containing 165 systems
    \item Z: Expert Score $<$ 0.8, containing 1208 systems
\end{itemize}
Figure \ref{Fig:categories_lenses} presents a random selection of eight systems from each category with their corresponding Expert Scores, providing an intuitive sense of the typical characteristics within each group.

\section{Results} \label{Sec:Results}

\subsection{Correlation Study}

To measure the agreement between two classifiers in their highest and lowest scoring predictions, we compute the Top-k Jaccard Similarity, defined as the Jaccard index between the sets of the top-k scored candidates for each classifier. For two top-k sets A and B, the Jaccard index is given by:
\begin{equation}
\text{Jaccard}(A, B) = \frac{|A \cap B|}{|A \cup B|}
\end{equation}
This metric ranges from 0 (no shared candidates) to 1 (identical top-k selections), and directly reflects practical alignment between classifiers in their high-confidence predictions. We adopt this metric to evaluate whether different classifiers prioritize the same candidates at the top (or bottom) of the ranking. Such an agreement would suggest that the classifiers are sensitive to similar morphological features and assign the highest or lowest scores to similar subjects.

\begin{figure*}[htbp]
 \centering
 \includegraphics[width=\textwidth]{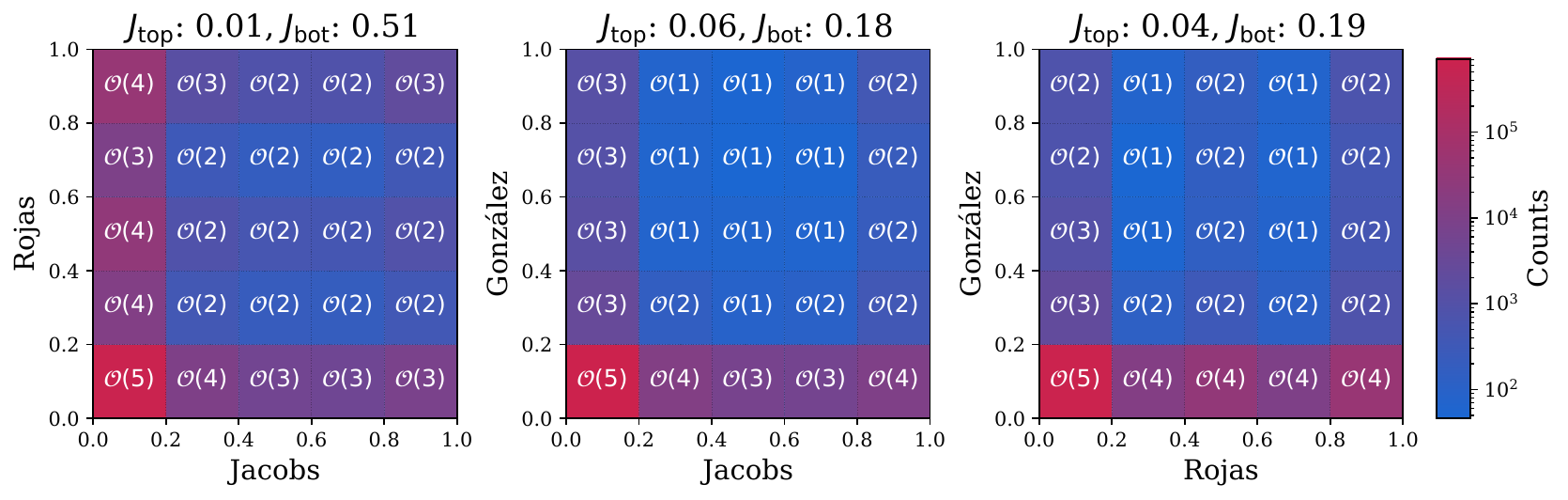}
\caption{2D histograms comparing ML normalized ranks from the three ML models for all subjects processed by the three ML models (Intersection sample). Each subfigure displays a different pairwise comparison, along with the corresponding Top-5000 and Bottom-100,000 Jaccard Similarity values.}
\label{Fig:2d_Histo_correlation_ml}
\end{figure*}

To illustrate the correlation between the ML models, we present Figure~\ref{Fig:2d_Histo_correlation_ml}, which shows 2D histograms of the normalized ranks assigned by each pair of models across the entire Intersection sample. Notably, for all ML models, over 80\% of the targets in this sample received normalized rank values close to zero ($<$~0.007). Conversely, only a small fraction of targets receive intermediate normalized rank values (0.3–0.7). Both observations are expected as ML classifiers tend to be over-confident by construction. In the same figure, we also report the Jaccard Similarity Index values for the top 5000 and bottom 100,000 ranked subjects for each pair of models. We find low similarity in the top subsets, with the highest Jaccard value being 0.06 between the Jacobs and González models. In contrast, the bottom-ranked subsets exhibit substantially higher agreement, with the Jacobs–Rojas pair reaching a Jaccard value of 0.51. These results suggest that while the models differ in the types of strong lensing morphologies they prioritize, they converge more strongly in their identification of high-confidence, likely negative cases.

\begin{figure*}[htbp]
 \centering
 \includegraphics[width=\textwidth]{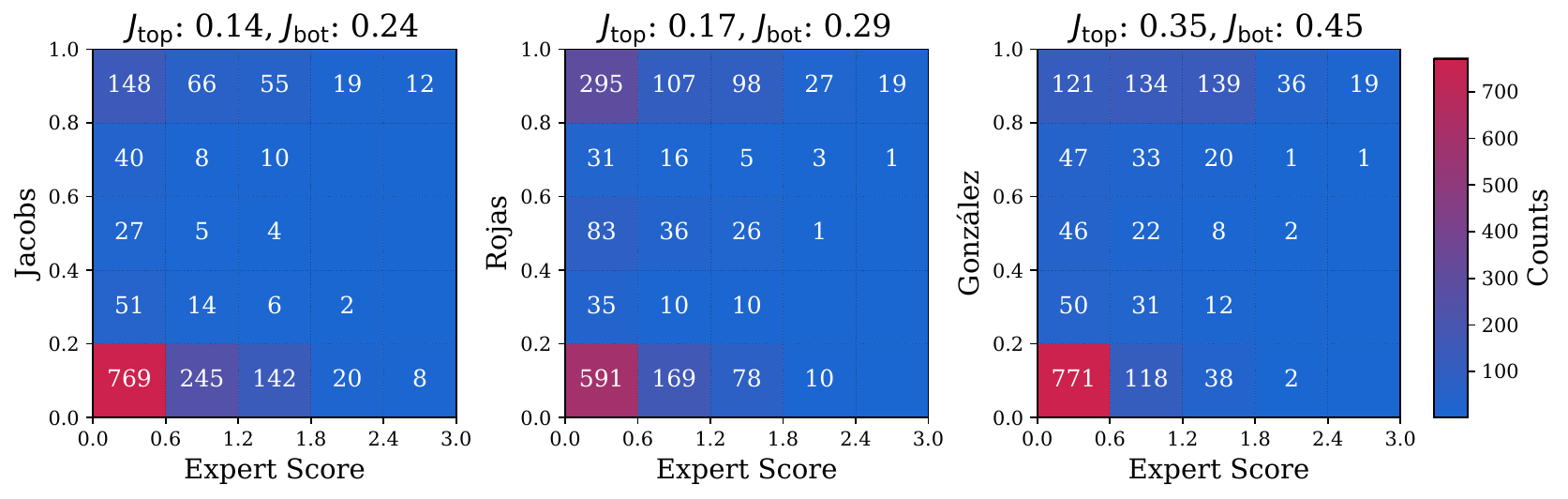}
\caption{2D histograms comparing Expert Scores with ML normalized ranks from the three ML models for the systems visually inspected by experts. The subtitles indicate the Jaccard Similarity values computed for the Top-200 and Bottom-600 ranked candidates.}
\label{Fig:2d_histo_ml_exp}
\end{figure*}

Next, we examine the correlation between the ML models and the expert scores. Figure~\ref{Fig:2d_histo_ml_exp} shows 2D histograms of the expert scores versus the normalized ranks from each ML model. This figure also reports the Jaccard Similarity Index values between the experts and each ML model for the top 200 and bottom 600 ranked subjects. We find that the agreement among the top-ranked subjects increases across successive ML models, rising from 0.14 for Jacobs to 0.35 for González. Similarly, the agreement among the bottom-ranked subsets is higher than for the top-ranked ones, and it also increases across successive ML models, from 0.24 for Jacobs to 0.45 for González. The significantly higher agreement between the experts and the González' model is likely due to their adoption of Interactive Machine Learning (see Sec \ref{subsec:method_gonzalez}) to construct the training sample. 

\begin{figure*}[htbp]
 \centering
 \includegraphics[width=\textwidth]{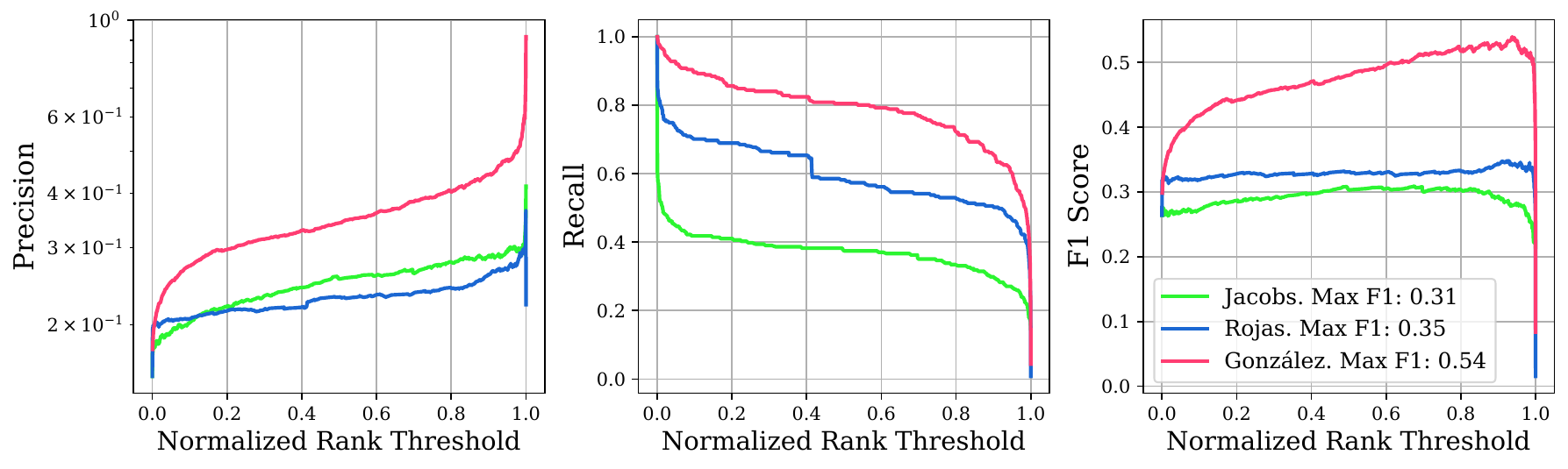}
\caption{Precision, recall, and F1-score as a function of normalized rank threshold for the three ML models. We use the sample of inspected systems to evaluate these metrics and assume a true strong lens is defined by an Expert Score~$>$~1.2. The last subfigure includes the maximum F1-score achieved by each model.}
\label{Fig:Performance_curves}
\end{figure*}

\begin{figure}[htbp]
 \centering
 \includegraphics[width=0.95\columnwidth]{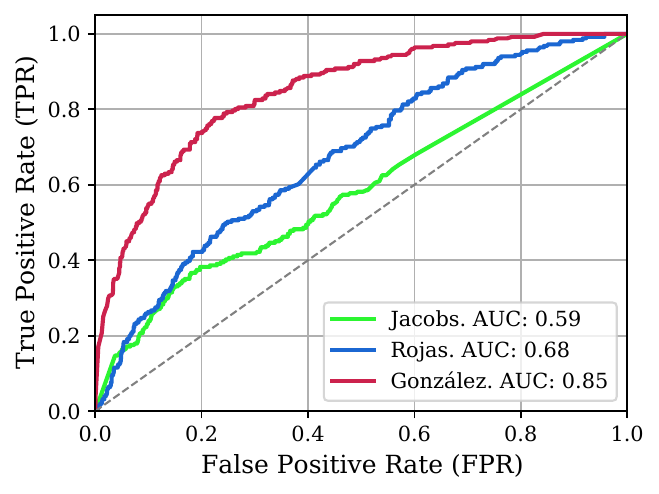}
\caption{Receiver Operating Characteristic (ROC) curves for the three ML models, evaluated on the sample of systems inspected by experts. A true strong lens is defined as a system with Expert Score~$>$~1.2. The legend includes the area under the curve (AUC) for each ML model. The FPR values reflect the much lower proportion of non-lenses in the inspected sample than in typical target search catalogs.}
\label{Fig:ROC_curve}
\end{figure}

\subsection{Performance Comparison}

Now, we evaluate the performance of the ML models using the standard metrics: recall, precision, F1-score, and area under the Receiver Operating Characteristic curve (AUROC). Recall or True Positive Rate (TPR) is often the equivalent of completeness and is defined as the proportion of all actual positives that were classified correctly as positives: $TPR = TP/(TP+FN)$. Precision, related to purity, is the proportion of all the model's positive classifications that are actually positive, and it is defined as $P=TP/(TP+FP)$. The F1-score is the harmonic mean of precision and recall, providing a single metric that balances both, and it's defined by the equation: 
\begin{equation}
    F_1 = \frac{2 \times \text{precision} \times \text{recall}}{\text{precision} + \text{recall}} \\
    = \frac{2TP}{2TP + FP + FN}.
\label{eq:f1_score}
\end{equation}
This value ranges from 0 to 1, with higher values indicating a better trade-off between recall and precision. A ROC curve illustrates an ML model's ability to distinguish between positive and negative examples by plotting the TPR against the False Positive Rate (FPR) for varying probability thresholds. The AUROC quantifies this ability, with values ranging from 0 (perfectly incorrect) to 1 (perfect), and 0.5 indicating random guessing.

To calculate these performance metrics, we consider the systems inspected by the experts and assume that all subjects with an Expert Score $\ge$ 1.2 (A and B categories of confidence) are true strong gravitational lenses (positives), while all others are considered not lenses (negatives). Figure~\ref{Fig:Performance_curves} shows how precision, recall, and F1-score vary as a function of the normalized rank threshold for each ML model. Among the three models, González's performs best overall, while Jacobs' achieves higher precision than Rojas', who, in turn, attains a significantly higher recall. In terms of F1-score, González achieves the highest value of 0.54, while Rojas and Jacobs obtain 0.35 and 0.31, respectively. Figure~\ref{Fig:ROC_curve} illustrates the ROC curves for the three ML models. The FPR values in the figure are influenced by the relatively small proportion of non-lenses in the inspected sample compared to the original target search catalogs. As a result, the FPRs shown here may underestimate the rate of false positives in a real-world application. In terms of AUROC, González achieves the highest value of 0.85, followed by Rojas and Jacobs with values of 0.68 and 0.59, respectively.

\subsection{Completeness Study}

\begin{figure*}[htbp]
 \centering
 \includegraphics[width=\textwidth]{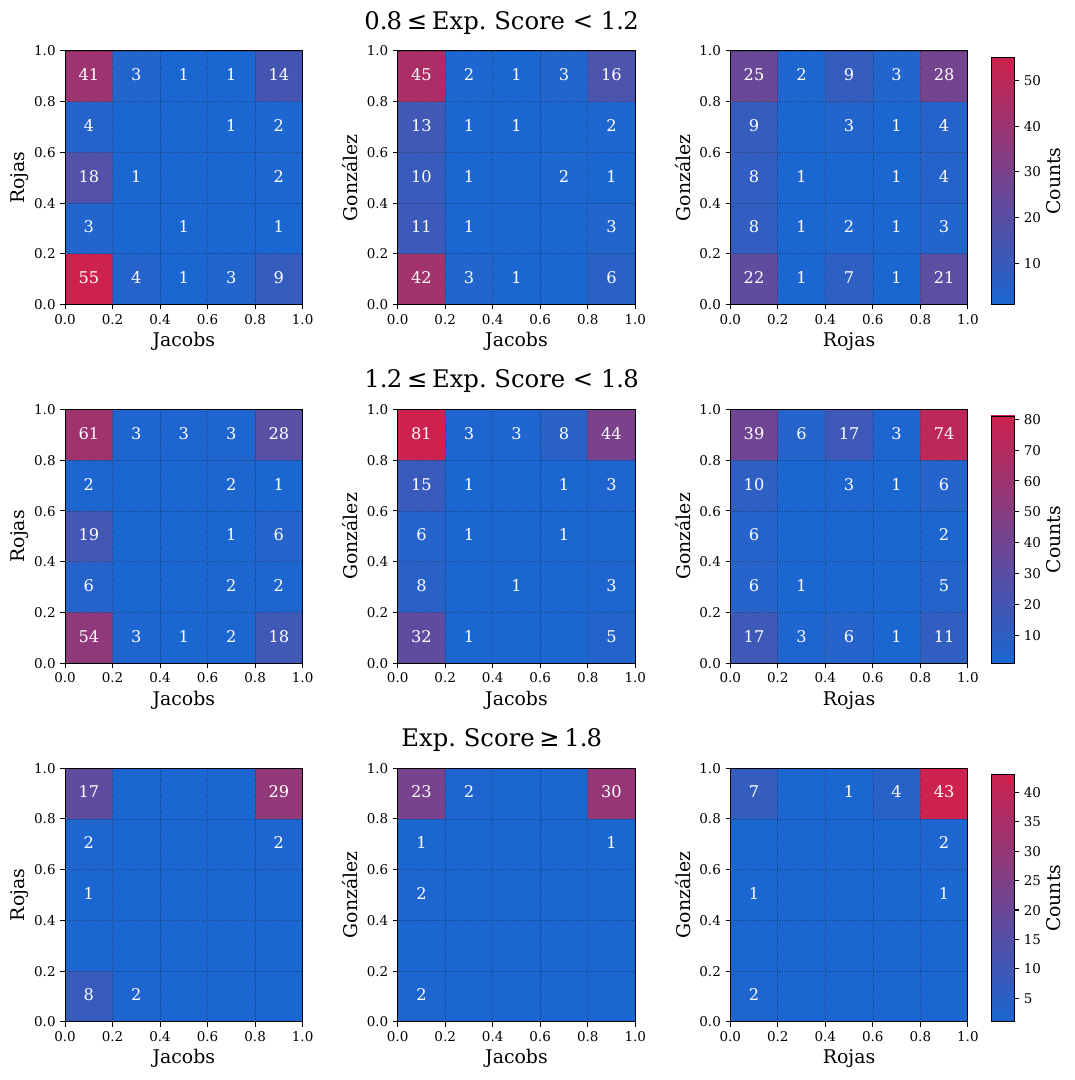}
\caption{2D histograms comparing ML normalized rank values from the three ML models for systems with Expert Score $\ge$ 0.8. Each row corresponds to a different confidence category, arranged from top to bottom as A, B, and C.}
\label{Fig:2d_histo_exp}
\end{figure*}

Figure~\ref{Fig:2d_histo_exp} presents 2D histograms comparing pairs of ML normalized ranks for strong lensing candidates grouped by different confidence categories. While about 36\% of the systems in the C category receive normalized rank values~$<$0.8 across all three ML models, only 22\% of those in the B category fall below this threshold. For candidates in the A category, the densest regions in the 2D histograms correspond to areas where at least one normalized rank value exceeds 0.8, with only 5\% of these high-confidence candidates receiving normalized ranks below 0.8 by the three networks. 


\begin{figure*}[htbp]
 \centering
 \includegraphics[width=\textwidth]{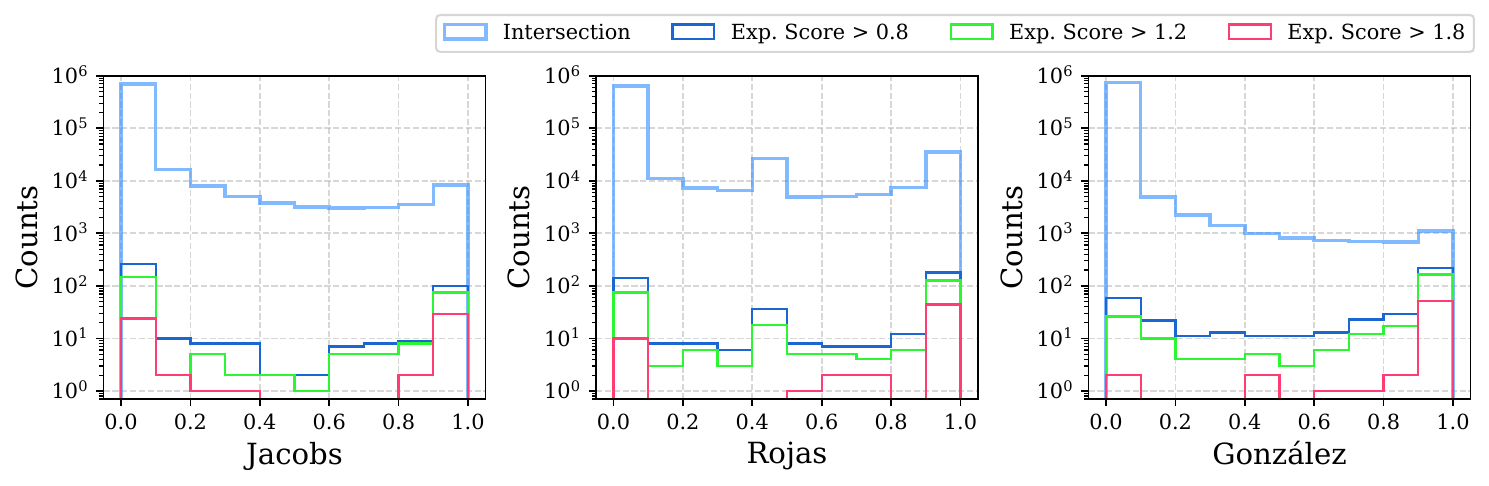}
\caption{Histograms of ML normalized rank values for each search effort. Each histogram includes all targets processed by all three searches (Intersection sample) and strong lensing candidates grouped by different confidence levels.}
\label{Fig:histo_good_cand}
\end{figure*}

We further analyze the distributions of normalized ranks assigned by each ML model to the systems visually inspected by experts and compare them to the overall distribution across the Intersection sample. Figure~\ref{Fig:histo_good_cand} presents histograms of ML normalized ranks for all targets in the Intersection sample, as well as for subsets of strong lensing candidates grouped by confidence level. As also illustrated by Figure~\ref{Fig:2d_histo_exp}, successive search efforts yield a higher fraction of moderate- to high-confidence candidates receiving high rank values: among candidates with an Expert Score $\geq$ 1.2, 31\%, 52\%, and 70\% received a normalized rank $\geq$ 0.8 from the Jacobs, Rojas, and González models, respectively. However, the exact reasons for varying ``completeness" rates among the three ML models are unclear due to the fundamental differences in the search methodologies. Additionally, the notice that the models are complementary: 82\% of the candidates with an Expert Score $\geq$ 1.2 are assigned a normalized rank $\geq$ 0.8 by at least one of the three classifiers, significantly higher than the best individual ML model.

Figure~\ref{Fig:histo_good_cand} also shows that, for all ML models, the fraction of strong lensing candidates with Expert Score~$\geq$~0.8 relative to the total number of targets drops sharply for normalized ranks below 0.8. This trend supports adopting high normalized rank thresholds when selecting systems for visual inspection, thereby improving efficiency by focusing human effort on the most promising candidates.

\begin{figure*}[htbp]
 \centering
 \includegraphics[width=\textwidth]{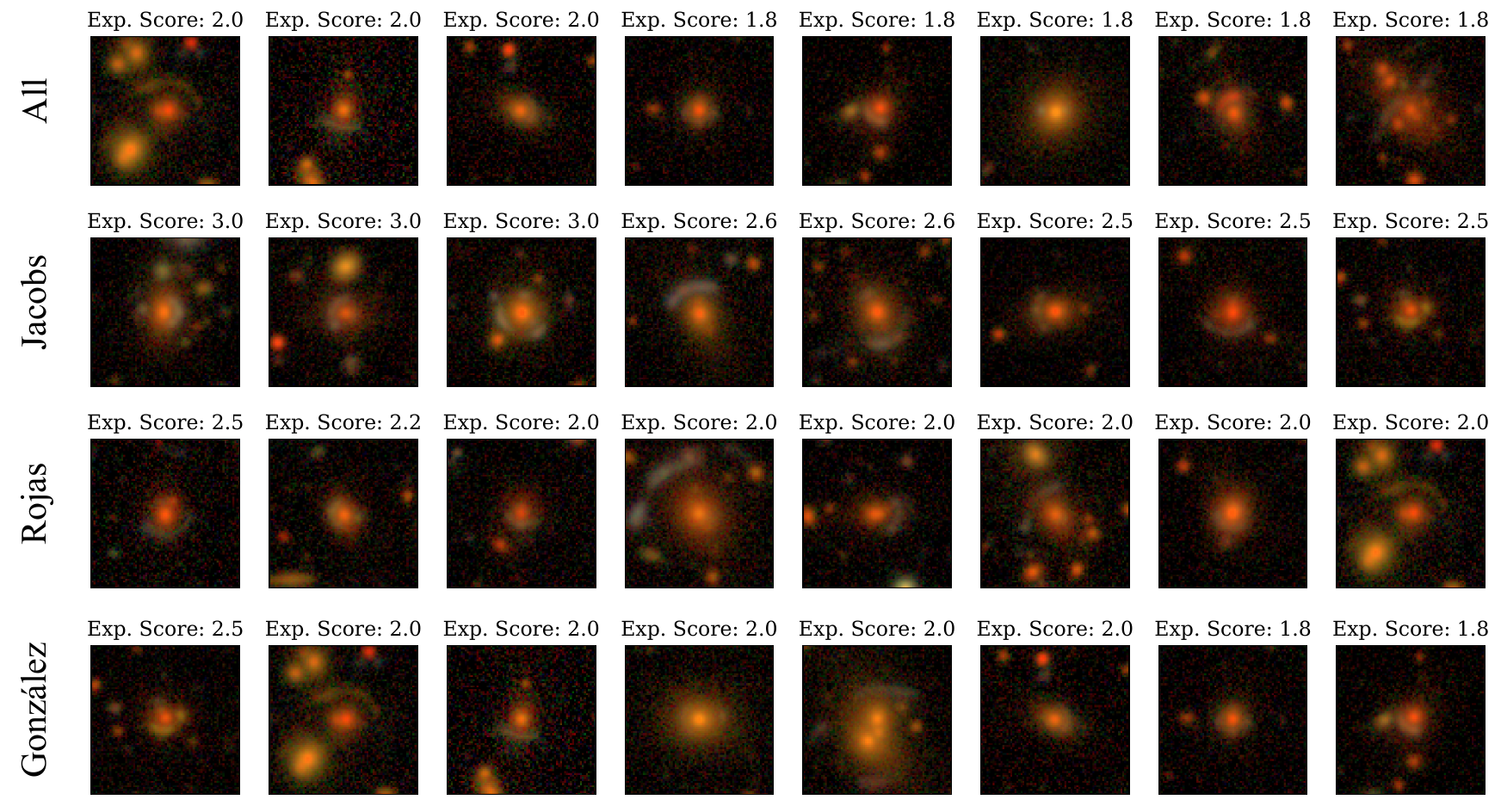}
\caption{Collage of strong lensing candidates with the highest Expert Scores among those assigned ML normalized ranks below 0.8. The first row shows candidates with normalized ranks~$<$0.8 across all three ML models, while the following rows correspond to those with normalized ranks~$<$0.8 in Jacobs, Rojas, and González, respectively. Within each row, candidates are sorted by Expert Score from highest to lowest, with the Expert Score displayed in the title of each image.}
\label{Fig:qualitative_figure}
\end{figure*}

As a qualitative analysis, Figure~\ref{Fig:qualitative_figure} presents a collage of the highest-confidence strong lensing candidates that received ML normalized ranks below 0.8 across all networks, and for each individual ML model. This figure helps identify potential common features among the ``missed" candidates, providing insight into whether the ML models fail to recover promising systems or if these cases are inherently challenging. The first row in this figure, which shows candidates with ML normalized ranks~$<$0.8 by all three networks, contains only three candidates within the A category, and although the first two examples have bright arclets, they appear somewhat flat. Most of the remaining candidates in this row exhibit lensing features that are faint, small, uncommon, or group scale. For individual works, it is difficult to pinpoint a consistent pattern among the ``missed" candidates. However, the row for Rojas appears to contain more systems with very small red-deflector galaxies and small Einstein radii, while the row for González primarily includes systems with flat or faint arclets, and small Einstein radii.

\section{Ensemble Methods}\label{sec:Ensemble_methods}

The ensemble techniques we investigate in this work are simple averaging, median, linear regression, a decision tree, a random forest, and an ``Independent Bayesian" method. In this section, we only consider the sample of systems visually inspected by experts, and this dataset is split into 80\% for training (and validation) and 20\% for testing. All techniques—except for averaging and median—are trained or fitted on the training sample. In this section, performance metrics are evaluated by defining a true strong lens as any system with an Expert Score~$\geq$~1.8 (i.e., belonging to the A confidence category); otherwise, the subject is treated as a negative example.

The hyperparameters of the decision tree and random forest classifiers were selected using a grid search combined with five‑fold stratified cross‑validation on the training sample. In stratified k‑fold cross‑validation, the data is partitioned into k subsets (here, k=5) that preserve the class distribution of the full dataset; the model is iteratively trained on k-1 of these subsets and validated on the remaining one, cycling through all folds. This approach allows all available training data to be used for both fitting and validation, which is especially important given the small fraction of true lenses in our sample. We explored a range of hyperparameter combinations for each model and selected those that maximized the F1 score, using class‑balanced weights to mitigate the effect of class imbalance. The models with the best-performing hyperparameters were finally trained on the complete training sample.

Each ensemble method receives as input the ML raw scores, as we noticed a consistent drop in performance across nearly all methods when using normalized ranks. The linear regression is fitted to predict the Expert Score, and, during testing, its predictions are divided by 3 to normalize the output. The decision tree and random forest models are trained as classification tasks with Boolean labels indicating whether a candidate is a genuine strong lens. During testing, we use the \texttt{predict\_proba()} function for these two models to obtain, for each input, the model‑estimated probability of being a lens (i.e., a value between 0 and 1). Consequently, the outputs of all ensemble methods are normalized and can be interpreted as an estimate of the confidence that a subject is a lens. These output values are then used to calculate performance metrics on the testing sample. 

The ``Independent Bayesian" method is inspired by the best-performing ensemble technique in \cite{Holloway_2024}: we first calibrate the ML raw scores from each individual model and then combine the calibrated scores using a Bayesian approach. The calibrated probabilities in the training sample are defined to yield statistically interpretable outputs: among all subjects assigned a calibrated probability of X, approximately X percent are actual strong lenses (again defined as A-category subjects). To obtain these calibrated scores, we fit an isotonic regression function to each ML model, mapping its original ML score to the corresponding calibrated probability. During testing, the three predicted calibrated scores are then combined using Bayes' Theorem, assuming that the outputs of the individual ML models are independent \citep[equation 11 from][]{Holloway_2024}.

We evaluate the performance of our ensemble methods using the testing sample. Given the large class imbalance in this dataset—only 12 positive examples out of \raisebox{0.5ex}{\texttildelow}330 subjects—we focus on the maximum F1 score and precision at a chosen completeness value. To assess the stability and uncertainty of each method, we apply bootstrapping: we generate multiple resampled versions of the testing set by sampling with replacement, and evaluate the performance on each new sample. This approach allows us to estimate both the average maximum F1 score and the associated uncertainty for each ensemble technique.

\begin{figure}[htbp]
 \centering
 \includegraphics[width=1\columnwidth]{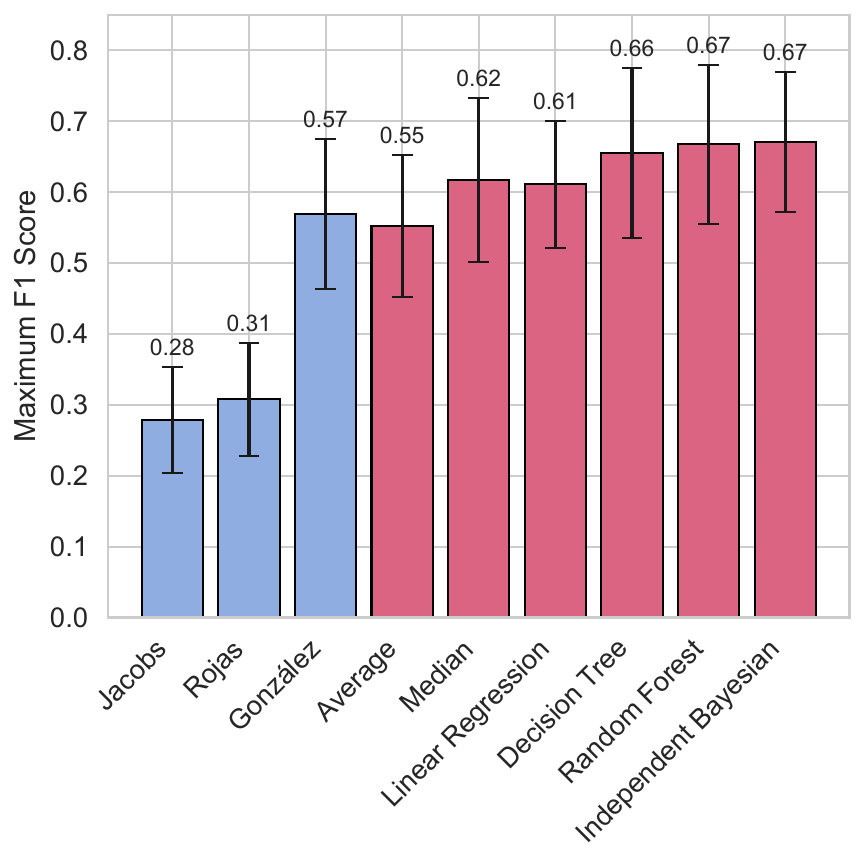}
\caption{Maximum F1 scores achieved by each ensemble method, evaluated on a test set derived from the expert-inspected sample. Bootstrapping is implemented to estimate mean performance and uncertainty. Results for the three individual ML models are shown for reference.}
\label{Fig:Ensemble_f1scores}
\end{figure}

Figure~\ref{Fig:Ensemble_f1scores} shows the maximum F1 score achieved on average by each ensemble method, along with the associated uncertainties. For reference, the figure also includes the corresponding values for the individual ML models. All ensemble methods—except for the simple average—outperform the individual models in terms of maximum F1 score. This finding is consistent with \citet{euclid_holloway}, which reported that simple averaging can sometimes underperform relative to individual models, and consistently performs worse than the "Bayesian Independent" method. The highest maximum F1 scores are reached by the decision tree, random forest, and Bayesian Independent methods with a similar value of approximately~0.67. These values are significantly higher than those achieved by any individual ML model.

Figure~\ref{Fig:Ensemble_f1scores} shows the maximum F1 score reached by each ensemble method on average, along with the associated uncertainties. The figure also includes the corresponding values for the individual ML models as reference points. All ensemble methods, except for the simple average, achieve a higher maximum F1 score than any individual ML model. \citep{euclid_holloway} also found that an average of individual ML model scores sometimes performs worse than individual models and that it always performs worse than the ``Bayesian Independent" method. According to this metric, the best-performing ensemble techniques are the decision tree, the random forest, and the Independent Bayesian method, which exhibit very similar performance with maximum F1 scores of \raisebox{0.5ex}{\texttildelow}0.67. The maximum F1 scores of these three methods are significantly higher than the best individual ML model (0.57).


As an additional performance comparison, we evaluated the precision achieved by each method at a chosen completeness of approximately two‑thirds (i.e., recovering 8 of the 12 true lenses in the testing set). For each method, we determined the score threshold corresponding to this completeness and counted the number of images in the intersection sample with scores above this threshold, from which we derived the resulting precision. Due to rounding the individual ML scores to four significant digits, several of the 8 candidates share identical scores, preventing us from selecting the same completeness value for all; we therefore report the closest possible values. The results are summarized in Table~\ref{tab:ensemble_precision}. The Independent Bayesian and linear regression methods achieved the highest precision values (13.0\% and 12.5\%, respectively), followed by the random forest (6.8\%), all of which exceed the precision from the best individual ML model (2.35\%). The remaining ensemble techniques performed worse than the top individual model. We emphasize, however, that these estimates are subject to large uncertainties due to the limited number of true lenses in the testing set, as also reflected in the wide bootstrap uncertainties for the maximum F1 scores in Figure~\ref{Fig:Ensemble_f1scores}.

\begin{table*}[htbp]
\centering
\begin{threeparttable}
\caption{Precision of each method at approximately two‑thirds completeness.}
\label{tab:ensemble_precision}
\begin{tabular}{lcccc}
\toprule
\textbf{Method} & \textbf{{\#} of lenses} & \textbf{Completeness (\%)} & \textbf{{\#} of images} & \textbf{Precision (\%)} \\
\midrule
Jacobs                & 8  & 66.67 & 86021 & 0.01 \\
Rojas                 & 8  & 66.67 & 19139 & 0.04 \\
González              & 10 & 83.33 & 426   & 2.35 \\
Average               & 8  & 66.67 & 873   & 0.92 \\
Median                & 8  & 66.67 & 699   & 1.14 \\
Linear Regression     & 10 & 83.33 & 80    & 12.50 \\
Decision Tree         & 8  & 66.67 & 293   & 2.73 \\
Random Forest         & 8  & 66.67 & 117   & 6.84 \\
Independent Bayesian  & 9  & 75.00 & 69    & 13.04 \\
\bottomrule
\end{tabular}
\end{threeparttable}
\end{table*}

\section{Conclusions} \label{Sec:Conclusions}

The main goals of this work are to evaluate the performance and completeness of different ML models in detecting strong gravitational lenses and to investigate ensemble techniques for constructing a more robust classifier. To achieve this, we compiled the ML scores from three independent search efforts applied to DES data \citep{Jacobs_2019_high, Jacobs_2019_extended, Rojas_2022, gonzalez2025}. From the full set of targets analyzed by all three works (the Intersection sample), we selected those reported as strong lensing candidates in the SLED database. A team of strong lensing experts visually inspected DES Y6 cutout images of these candidates, classifying each into one of four confidence categories. We assigned a numerical value in the range 0-3 to these categories, with higher values indicating greater confidence in the candidate. We calibrated these expert assessments to obtain a final “Expert Score,” which we use to classify the inspected systems into four confidence categories: A (definite lenses), B (probable lenses), C (ambiguous cases), and Z (likely non‑lenses).

First, we analyzed the agreement between the ML models, as well as between each model and the team of experts. Using the Intersection sample, we found that the overlap among the top-ranked predictions is minimal, suggesting that the classifiers prioritize different lensing morphologies. In contrast, agreement is substantially higher for low-ranked targets: for example, the Jacobs and Rojas models share half of their bottom 100,000 ranked subjects. When comparing the ML predictions to expert assessments, we also observe stronger agreement in the bottom-ranked subsets, indicating greater consensus on what constitutes a clear non-lens than on what defines a high-confidence strong lens. Notably, the agreement between the expert scores and the ML predictions increases across successive search efforts, both for top- and bottom-ranked subjects. The highest agreement is observed for the González model, which likely reflects the benefits of incorporating Interactive Machine Learning (IML) to iteratively refine the training sample.

In terms of overall metrics, González’s model achieves the highest AUROC (0.85) and F1-score (0.54), outperforming Rojas (0.68, 0.35) and Jacobs (0.59, 0.31). These results highlight the steady improvement of ML-based strong lensing searches over time, suggesting that refinements in methodology, such as advancements in ML architectures and training data, contribute to enhanced performance. Unfortunately, pinpointing the exact reasons for one ML model outperforming another is not possible due to the substantial differences in their methodologies.

Our findings emphasize the effectiveness of ML techniques in identifying strong gravitational lenses, with only 5\% of the A‑category candidates receiving ML normalized ranks below 0.8 from all three networks. When evaluated on A‑ and B‑category candidates, each successive work attains a higher completeness rate, with values of 31\%, 52\%, and 70\% for the Jacobs, Rojas, and González models, respectively. Moreover, 82\% of the candidates in these two categories receive a normalized rank above 0.8 from at least one of the three ML models, a value substantially higher than the completeness achieved by any individual model alone. The complementarity of different ML models highlights the potential of leveraging diverse ML approaches and adopting high probability thresholds to maximize sample completeness while minimizing false positives, thereby optimizing human resources for visual inspection. Notably, our results also indicate that using simulated strong lenses with unrealistic features to train ML models does not necessarily hinder performance. This is evidenced by González’s model, which, although trained on such simulations, achieves the highest recovery of strong lensing candidates.

We also explored the effectiveness of ensemble methods that combine the outputs of the three individual ML models. The techniques tested include simple averaging, median, linear regression, decision tree, random forest, and an Independent Bayesian method. Except for the simple average, all ensemble methods achieved higher maximum F1 scores than the best individual ML model, with the decision tree, random forest, and Independent Bayesian approaches performing best. As a complementary evaluation, we also measured precision at a fixed completeness of approximately two‑thirds, finding that the Independent Bayesian, linear regression, and random forest methods provided the highest precision (13.0\%, 12.5\%, and 6.8\%, respectively), substantially outperforming the best individual ML model (2.4\%). These results highlight the potential of ensemble techniques to drastically reduce the number of false positives in strong lensing searches. A similar conclusion was reached by \citet{euclid_holloway}, who found that combining multiple ML networks with citizen science scores can maintain high completeness while significantly reducing false positives. Further work is needed to understand how ML models can best leverage human-in-the-loop approaches and crowd-sourced input to optimize human effort in the upcoming era of big-data astronomy and strong-lens discovery.

\section*{Acknowledgements}

This research made use of the services provided by the OSG Consortium \citep{osg_06, osg07, osg09, osg_2015}, which is supported by the National Science Foundation awards \#2030508 and \#1836650.

JG was supported by the U.S. Department of Energy through grant DE-SC0017647.

KB has been supported by the U.S. Department of Energy Early Career Program through grant DE-SC0022950.

This project has received funding from the European Research Council (ERC) under the European Union’s Horizon 2020 research and innovation programme (LensEra: grant agreement No 945536).

T.~E.~C. is funded by a Royal Society University Research Fellowship. 

DS acknowledges the support of the Fonds de la Recherche Scientifique-FNRS, Belgium, under grant No. 4.4503.1.

PH acknowledges funding from the Science and Technology Facilities Council, Grant Code ST/W507726/1





\appendix
\section{Origin of Candidates}\label{app:sled}

Table~\ref{tab:sled_sample} lists the publications that report the 1651 systems in the sample visually inspected by experts. The table indicates the astronomical survey each paper is based on and the number of candidates overlapping with the inspected sample. Some individual systems are reported in more than one publication. Survey abbreviations are as follows: DES = Dark Energy Survey, DESI-LS = Dark Energy Survey Instrument Legacy Imaging Surveys, GALAH = GALactic Archaeology with HERMES, GAMA = Galaxy And Mass Assembly, HSC = Hyper Suprime Camera, HST = Hubble Space Telescope, KiDS = Kilo-Degree Survey, Pan-STARRS = Panoramic Survey Telescope and Rapid Response System, SDSS = Sloan Digital Sky Survey, UNIONS = Ultraviolet Near Infrared Optical Northern Survey, and VOICE = VST Optical Imaging of the CDFS and ES1.
 
\begin{table}[ht]
    \centering
    \caption{Publications reporting the systems in the sample visually inspected by strong lensing experts}
    \begin{tabular}{l l r}
        \textbf{Reference} & \textbf{Survey} & \textbf{Count} \\
        \hline
        \cite{Jacobs_2019_extended}&DES &602\\
        \cite{Rojas_2022}&DES &241\\
        \cite{Diehl_2017}&DES &79\\
        \cite{Jacobs_2019_high}&DES &64\\
        \cite{ODonell_2022}&DES &43\\
        \cite{Nord_2020}&DES &8\\
        \cite{Agnello_2019}&DES &3\\
        \cite{Nord_2016}&DES &1\\
        \cite{Lemon_2020}&DES &1\\
        \cite{Storfer_2024}&DESI-LS&319\\
        \cite{Huang_2021}&DESI-LS&136\\
        \cite{Stein_2022}&DESI-LS&95\\
        \cite{Huang_2020}&DESI-LS&62\\
        \cite{He_2023}&DESI-LS&1\\
        \cite{Zhang_2023}&DESI-LS&1\\
        \cite{Ishigaki_2021}&GALAH \& Gaia&1\\
        \cite{Holwerda_2015}&GAMA&4\\
        \cite{Lemon_2022}&Gaia&1\\
        \cite{Shu_2022}&HSC&74\\
        \cite{Jaelani_2020}&HSC&54\\
        \cite{Ca_ameras_2021}&HSC&49\\
        \cite{Sonnenfeld_2017}&HSC&45\\
        \cite{Jaelani_2020a}&HSC&27\\
        \cite{Wong_2018}&HSC&21\\
        \cite{Chan_2020}&HSC&3\\
        \cite{Sonnenfeld_2019}&HSC&2\\
        \cite{Desprez_2018}&HST&1\\
        \cite{Petrillo_2019}&KiDS&212\\
        \cite{Li_2020}&KiDS&11\\
        \cite{Li_2021}&KiDS&9\\
        \cite{Ca_ameras_2020}&Pan-STARRS&20\\
        \cite{Talbot_2021}&SDSS&37\\
        \cite{Cao_2020}&SDSS&12\\
        \cite{Gavazzi_2014}&UNIONS&50\\
        \cite{More_2012}&UNIONS&12\\
        \cite{More_2015}&UNIONS&6\\
        \cite{Sonnenfeld_2013}&UNIONS&5\\
        \cite{Paraficz_2016}&UNIONS&4\\
        \cite{Cabanac_2006}&UNIONS&3\\
        \cite{Sygnet_2010}&UNIONS&1\\
        \cite{Jacobs_2017}&UNIONS&1\\
        \cite{Gentile_2021}&VOICE&2\\
        \hline
    \end{tabular}
    \label{tab:sled_sample}
\end{table}


%



\bibliography{sample631}{}
\bibliographystyle{aasjournal}



\end{document}